\documentclass[fleqn,usenatbib]{mnras}

\usepackage{newtxtext,newtxmath}

\usepackage[T1]{fontenc}
\usepackage{siunitx}

\DeclareRobustCommand{\VAN}[3]{#2}
\let\VANthebibliography\thebibliography
\def\thebibliography{\DeclareRobustCommand{\VAN}[3]{##3}\VANthebibliography}

\usepackage{graphicx}	
\usepackage{amsmath}	

\usepackage{scalerel,tikz}
\usetikzlibrary{svg.path}
\definecolor{orcidlogocol}{HTML}{A6CE39}
\tikzset{orcidlogo/.pic={
 \fill[orcidlogocol] svg{M256,128c0,70.7-57.3,128-128,128C57.3,256,0,198.7,0,128C0,57.3,57.3,0,128,0C198.7,0,256,57.3,256,128z};
 \fill[white] svg{M86.3,186.2H70.9V79.1h15.4v48.4V186.2z}
 svg{M108.9,79.1h41.6c39.6,0,57,28.3,57,53.6c0,27.5-21.5,53.6-56.8,53.6h-41.8V79.1z M124.3,172.4h24.5c34.9,0,42.9-26.5,42.9-39.7c0-21.5-13.7-39.7-43.7-39.7h-23.7V172.4z}
 svg{M88.7,56.8c0,5.5-4.5,10.1-10.1,10.1c-5.6,0-10.1-4.6-10.1-10.1c0-5.6,4.5-10.1,10.1-10.1C84.2,46.7,88.7,51.3,88.7,56.8z};
}}
\newcommand\orcidicon[1]{\href{https://orcid.org/#1}{\mbox{\scalerel*{
\begin{tikzpicture}[yscale=-1,transform shape]
\pic{orcidlogo};
\end{tikzpicture}
}{|}}}}

\title[Binary dwarf disruption in Sgr]{Hints of a disrupted binary dwarf galaxy in the Sagittarius stream}

\author[E. Y. Davies et al.]{
Elliot Y. Davies~\orcidicon{0000-0001-5996-4072}$^{1}$\thanks{E-mail: eyd20@cam.ac.uk},
Stephanie Monty~\orcidicon{0000-0002-9225-5822}$^{1}$,
Vasily Belokurov~\orcidicon{0000-0002-0038-9584}$^{1}$ and Adam M. Dillamore~\orcidicon{0000-0003-0807-5261}$^{1}$.
\\
$^{1}$Institute of Astronomy, University of Cambridge, Madingley Road, Cambridge CB3 0HA, UK}

\date{Accepted XXX. Received YYY; in original form ZZZ}

\pubyear{2015}

\begin{document}
\label{firstpage}
\pagerange{\pageref{firstpage}--\pageref{lastpage}}
\maketitle

\begin{abstract}
In this work, we look for evidence of a non-unity mass ratio binary dwarf galaxy merger in the Sagittarius stream. Simulations of such a merger show that, upon merging with a host, particles from the less-massive galaxy will often mostly be found in the extended stream and less-so in the central remnant.
Motivated by these simulations, we use APOGEE DR17 chemical data from approximately $1100$ stars in both the Sagittarius remnant and stream to look for evidence of contamination from a second dwarf galaxy. 
This search is initially justified by the idea that disrupted binary dwarf galaxies provide a possible explanation of the Sagittarius bifurcation, and the location of the massive, chemically peculiar globular cluster NGC 2419 found within the stream of Sagittarius. 
We separate the Sagittarius data into its remnant and stream and compare the [Mg/Fe] content of the two populations. In particular, we select [Mg/Fe] to search for hints of unique star formation histories among our sample stars.
Comparing the stream and remnant populations, we find regions have distinct [Mg/Fe] distributions for fixed [Fe/H], in addition to distinct chemical tracks in [Mg/Fe] -- [Fe/H] abundance space. 
We show that there are large regions of the tracks for which the probability of the two samples being drawn from the same distribution is very low ($p < 0.05$). Furthermore, we show that the two tracks can be fit with unique star formation histories using simple, one zone galactic chemical evolution models.
While more work must be done to discern whether the hypothesis presented here is true, our work hints at the possibility that Sagittarius may consist of two dwarf galaxy progenitors.
\end{abstract}




\section{Introduction}

A vital process in the growth of a typical large galaxy is the hierarchical accretion of smaller galaxies over time \citep[][]{white1978core}. The most striking example of this so-called \textit{galactic cannibalism} is currently underway quite close to home; the Sagittarius dwarf spheroidal galaxy (Sgr) is a nearby ancient disrupting satellite whose tidal stream encircles the Milky Way (MW) \citep[][]{ibata1995sgr, majewski2003two}. While the Sgr core is only 16 kpc from the Galactic centre, its stellar debris can be found over $100$ kpc away \citep[e.g.][]{newberg2003sagittarius, ruhland2011structure, belokurov2014precession, sesar2017distant}. Before its Galactic infall, Sgr is thought to have had a total mass of around $4\times10^{10}$ M$_{\odot}$ or greater \citep[e.g.][]{yanny2009tracing, niederste-ostholt2010re,  read2019sgr, dillamore2022impact}. Such a massive dwarf galaxy (dGal) will naturally bring a variety of substructure with it, including globular clusters (GCs). Given Sgr's vast coverage, we expect to find these associated GCs scattered throughout the Galaxy, as well as tightly bound to the Sgr core. Within the central core (or \textit{remnant}), we find many GCs like M~54, Ter~7, Ter~8 and Arp~2. In addition to these, there are now several GCs that have been physically and kinematically confirmed to belong to the extended stream of Sgr. These include Pal~12, Whiting~1, NGC~4147, NGC~5634 and NGC~2419 \citep[][]{bellazzini2020globular}.

Distinct among the Sgr-associated GCs is the exceptionally peculiar NGC~2419. Specifically, NGC~2419 is among the ten most massive GCs, with an estimated mass of around $10^6$ M$_{\odot}$ \citep[][]{Baumgardt2018}. Moreover, it has the highest spread in Mg of any known GC \citep[][]{cohen2011peculiar, mucciarelli2012news}. While much evidence points towards it being a Sgr member GC, it is one of the furthest known MW clusters at a Galactocentric radius of approximately 90 kpc \citep[][]{harris2010new, baumgardt2021accurate}, and has one of the longest relaxation times of any similar object \citep[$\sim10$~Gyr according to][]{Baumgardt2018}. Like most GCs, NGC 2419 is thought to be very deficient in dark matter (DM). \citet[][]{ibata2013do} show that, while the presence of DM cannot be ruled out, it is likely to be highly centrally concentrated if it is present at all. Moreover, \citet[][]{bellazzini2007surface} note that the half light radius of NGC~2419 is approximately 5 times that of a typical GCs of a similar luminosity, which is not an unreasonable size for a nucleus of a dGal or an ultra compact dwarf \citep[e.g.][]{federici2007extended, Mackey2005}. With these properties in mind, it has been proposed that NGC~2419 may in fact be the nuclear star cluster of a now-disrupted dGal \citep[e.g.][]{vandenbergh2004globular, Mackey2005, cohen2011peculiar, mucciarelli2012news, pfeffer2021accreted, davies2023disrupted}.

While the details of Sgr's past evolution are steadily being uncovered, many of its properties still remain a mystery. In particular, the stream has an unexplained bifurcation, most notably in the leading tail \citep[][]{belokurov2006field}, but also in the trailing tail \citep[][]{Koposov2012}. Naively, we would expect that a simple disrupted spheroidal system would form non-bifurcated tidal tails. The first explanation was put forward by \citet[][]{Fellhauer2006}, who suggested that the bifurcation resulted from two different wraps of the stream. This original hypothesis was ruled out by the similar properties -- distances, velocities, metallicities -- of both arms \citep[][]{yanny2009tracing, niederste-ostholt2010re}. Another explanation was put forward by \citet[][]{penarrubia2010was}. They suggested that the bifurcation could result if Sgr had a rotating disk. However, a lack of evidence for any internal rotation in the progenitor disfavours this scenario \citep[e.g.][]{penarrubia2011no}. So, while many attempts have been made to explain the bifurcation, there remains plenty of room for speculation. 

In a previous work, we introduced the idea that a disrupted binary dGal may be able to explain the bifurcation of Sgr and the location and size of NGC 2419 simultaneously \citep[][]{davies2023disrupted}. Specifically, we suggested the following: instead of Sgr being a single spheroidal dGal that disrupted in the MW, Sgr may have comprised two dGals that fully merged before this new coalesced system subsequently disrupts in a MW-like host potential. In this model, where we have created a new merged system with a now non-isotropic velocity distribution, the smaller dGal will have provided a sizeable population of eccentric stars that are tidally stripped by the MW-like host more easily than the stars of the larger dGal. Moreover, stars in the larger dGal are also likely to be heated from the initial merger. While the parameter space of the initial conditions for such a scenario remains wide open, our simulations show that the smaller galaxy will often have much more material in the stream upon disruption in the MW-like host. 

In the true spirit of \textit{Galactic Archaeology}, we turn to the chemical composition of stars in the Sgr stream in an attempt to untangle any possibly intertwined dynamical properties. If Sgr were to be composed of two disrupted dGals instead of one, we should see this imprinted in the chemical evolution of Sgr. According to our model, in which the smaller satellite's stars are found much more in the stream than the remnant (after disruption), we may expect to see a difference in the chemical abundances of stars throughout the stream. We expect this to arise given the unique star formation histories of the two galaxies prior to the merger. The evolution of light elements, or $\alpha$-elements, as a function of metallicity ([Fe/H]) has been shown to trace star formation histories \citep{tinsley1979}. Therefore, we use a sample of publicly available known Sgr member stars, and their APOGEE DR17 chemical information, to explore the composition of these two components. 

In this work, we explore the $\alpha$-elements by focusing entirely on Mg, which has been shown to be a very pure $\alpha$-element that traces star formation history (SFH) very well \citep[][]{kobayashi2020origin}. Moreover, Mg is one of the most accurate $\alpha$-elements measured by APOGEE \citep[][]{Jonsson2018APOGEE}. We leave the study of other elements for a different work, but note that similar trends in some other elements appear to exist in our data set. While currently limited by small number statistics from our admittedly restricted sample size, we hope that that these results open further discussion into the possibility that the make-up of Sgr may be a composite system containing stars formed in 2 (or more) individual galaxies.

The outline of this work is as follows. In Sec.~\ref{sec:simulations}, we re-introduce the binary dGal scenario as described in \citet[][]{davies2023disrupted}, and discuss the distribution of the debris. We then describe the observational data used in Sec.~\ref{sec:data}. In Sec.~\ref{sec:results} we analyse the data and present out findings. Lastly, in Sec.~\ref{sec:conclusions} we summarise our results and draw conclusions.

\section{Simulations}\label{sec:simulations}

In this section, we give an overview of the details of our simulations, which are of the same nature as the simulations described in \citet[][]{davies2023disrupted}. The $N$-body simulations snapshots presented in this work are intended to justify our choice of splitting the observational data into the \textit{stream} and \textit{remnant} components of Sgr. As before, a suite of $~100$ simulations were run, yet we only present a small sample of them here. The simulations, which involve the disruption of a binary dGal satellite system, are split into two parts. The first part -- the \textit{pre-infall} merger -- refers to the merging of the two satellites in the absence of a host (the MW). The second part -- the \textit{post-infall} merger -- refers to the merging of the resulting gravitationally bound system of the pre-infall merger with a static host MW potential.

\subsection{Pre-infall merger}

We begin by initialising two satellites of mass ratio $1:3$, with the smaller satellite placed at the virial radius of the larger. These satellites' potentials consist of spheroidal \citet[][]{hernquist1990analytical} models, and we use isotropic distribution functions established by the Eddington inversion formula \citep[Eq.~4.46b in][]{binneyandtremaine2008}. The larger satellite has a total mass of $3\times10^{10}$ M$_{\odot}$, and consists of $9\times10^5$ particles. The smaller satellite has a total mass of $1\times10^{10}$ M$_{\odot}$, and consists of $3\times10^5$ particles. These masses are chosen to total $4\times10^{10}$ M$_{\odot}$, which is a reasonable mass for Sgr according to e.g. \citet[][]{niederste-ostholt2010re, bennett2022exploring}. All properties of the pre-infall merger are fixed, from simulation to simulation, except the initial orbital circularity, $\eta = L/L_{\rm circ}(E)$, i.e. the ratio of the instantaneous angular momentum to the angular momentum of a circular orbit with the same energy. All simulations presented here have circularities from $\eta = 0.2$ to $\eta = 0.6$ to cover a sensible range. The satellites are evolved using the python package \textsc{pyfalcON}, a stripped down python interface of the \textsc{gyrfalcON} code \citep[][]{dehnen2000celestial}. Once the satellites have merged for six~Gyr, so that they form a fully merged and new gravitationally bound system, we save the positions and velocities of their constituent particles.

\subsection{Post-infall merger}

We place the resulting system of the pre-infall merger at $(x,y,z) = (0,0,80)$ kpc above the centre of the plane of a static host potential, with initial velocity $(v_x, v_y, v_z) = (108, 64, -128)$ km/s, to produce an approximately Sgr-like polar orbit with initial apocentre $r_{\rm apo} = 30$ kpc and pericentre $r_{\rm peri} = 90$ kpc. After approximately, 4 Gyr, the pericentre and apocentre of the system are $r_{\rm peri} = 20$ kpc and $r_{\rm peri} = 90$ kpc respectively. For this post-infall merger, we use the realistic MW-like fixed profile \textsc{MilkyWayPotential} from \textsc{Gala} \citep[][]{price-whelan2017gala}. The pre-infall system remains as a self-gravitating collection of $N$-body particles. It is important to note that the orbit of the post-infall merger is not intended to reproduce the exact orbital properties of Sgr, but instead simply follow a roughly Sgr-like orbit. The orientation of the plane of interaction of the initial pre-infall merger is varied with respect to the plane of the MW-like host (i.e. the MW disk plane) from simulation to simulation. This means that while the orbit of the post-infall merger is always the same, the different initial orientations of the pre-infall system allow us to achieve a variety of results.

\subsection{Distribution of post-infall debris}\label{sec:debris_distribution}

\begin{figure*}
    \centering
    \includegraphics[width=0.8\textwidth]{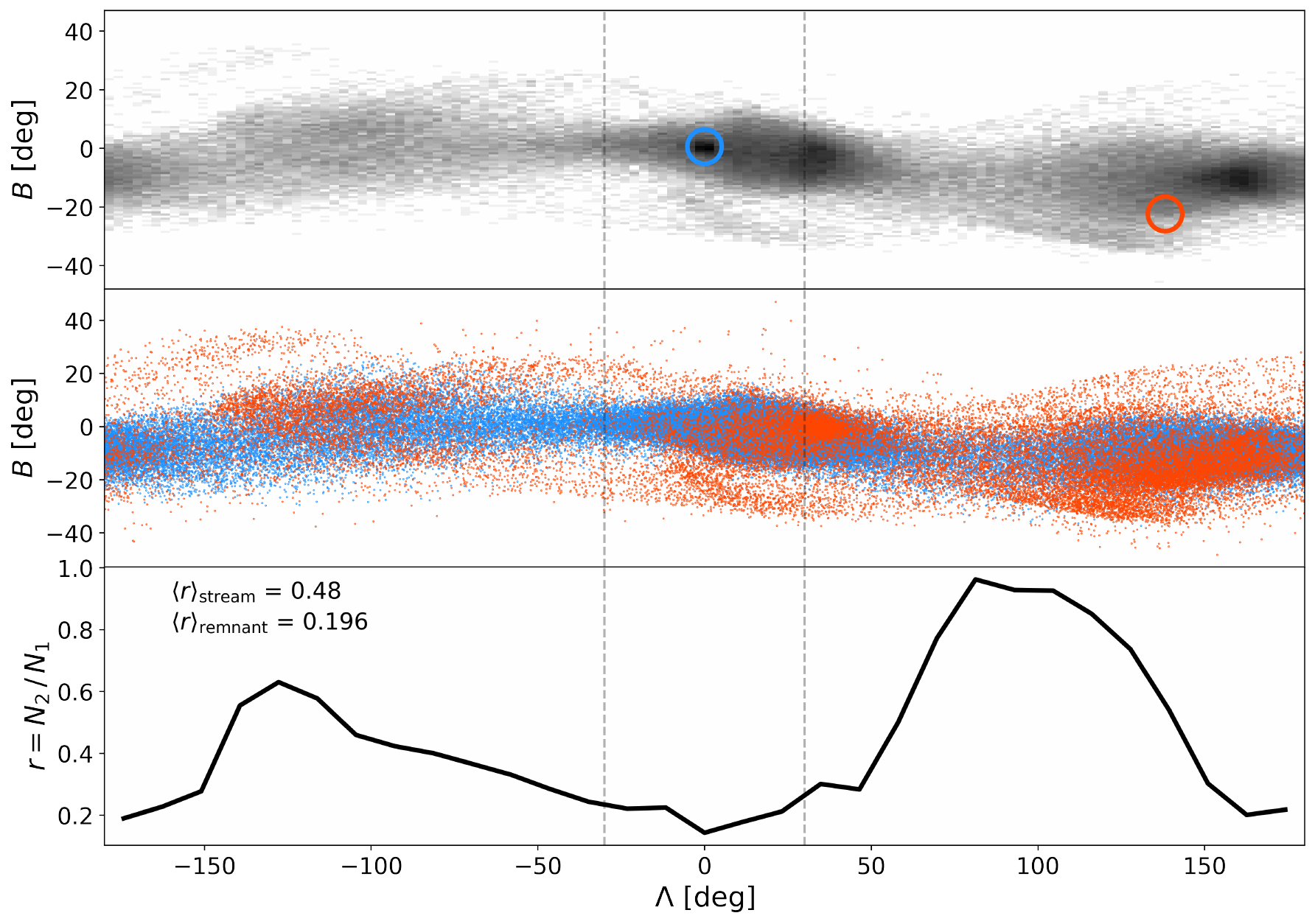}
    \caption{Snapshot of disrupted binary dGal simulation, after 4 Gyr of evolution in host potential, shown in stream coordinates $(\Lambda, B)$. In every panel of the figure, orange colours are to be associated with the smaller satellite, and blue with the larger satellite. \textit{Top:} the log density distribution of all particles in the simulation. The blue and orange circles illustrate the positions of the most bound particles of the satellites. \textit{Middle:} scatter plot of the particles of the two satellites. \textit{Bottom:} ratio of particles in the smaller satellite to the larger satellite, binned by stream coordinate $\Lambda$. We can see that the smaller satellite's contribution to the satellite is much larger in the stream than in the central remnant (i.e. $|\Lambda| \lesssim 50$ deg).}
    \label{fig:LambdaBratio}
\end{figure*}

We find that a common occurrence of our simulation is the following: the ratio $r = N_2 / N_1$ is higher in the stream than in the remnant after significant disruption in the host (i.e. several Gyr into the post-infall merger), where $N_2$ is the number of particles in the smaller satellite and $N_1$ is the number of particles in the larger satellite. In essence, the contamination from a second satellite is likely to be much higher in the stream than the remnant.

To illustrate this, we present an example of disrupted binary dGals in Fig.~\ref{fig:LambdaBratio}. In this example, the smaller satellite in the pre-infall merger had an initial orbital circularity of $\eta = 0.6$, and the plane of interaction of the pre-infall merger is parallel to the $z=0$ plane of the MW-like host. In the figure we present the debris stream-coordinates $(\Lambda, B)$, 4~Gyr into the post-infall merger with the host, with $\Lambda$ increasing towards the leading tail. These stream-coordinates were obtained by a great circle transformation. The origin of the great circle was selected as the most on-sky position of the bound particle $(\alpha, \delta) =(\alpha_0, \delta_0)$ (at the core of the remnant). The pole was calculated by first fitting a polynomial to the debris in on-sky coordinates and then selecting points at $\alpha_0 \pm 45$ degrees along the polynomial. The pole is then the normal to the plane defined by those points.

In the top panel of Fig.~\ref{fig:LambdaBratio}, we plot the logarithm of 2D density of debris particles in the stream coordinates, with the positions of the most bound particles of the two satellites highlighted. The blue circle shows the location of the larger satellite's most bound particle, and the orange circle shows the location of the smaller satellite's most bound particle. Under the blue circle, we see a high-density region corresponding to the location of the core of the larger satellite. The distant location of the orange circle demonstrates that the core of the smaller satellite has been ejected from the larger satellite. Additional high-density regions seen at $\Lambda \sim 30^\circ$ and $\Lambda \sim 160^\circ$ are apocentric pile-ups. In the middle panel of Fig.~\ref{fig:LambdaBratio} we show a scatter plot detailing the distribution of the debris from the two satellites. The orange scatter points are the particles from the smaller satellites and the blue are from the larger satellite. It is clear to see how the smaller satellite creates interesting bifurcation-like features at the edge of the stream, which are also evident in the top panel. In the bottom panel we plot the ratio $r = N_2 / N_1$ as a function of $\Lambda$. In this example, the smaller satellite's particles clearly contaminate the outer stream much more than around the central remnant. The more eccentric and high energy orbits of the particles in the smaller satellite likely make them more prone to being tidally pulled away from the centre of the pre-infall remnant.

\begin{figure}
    \centering
    \includegraphics[width=0.9\columnwidth]{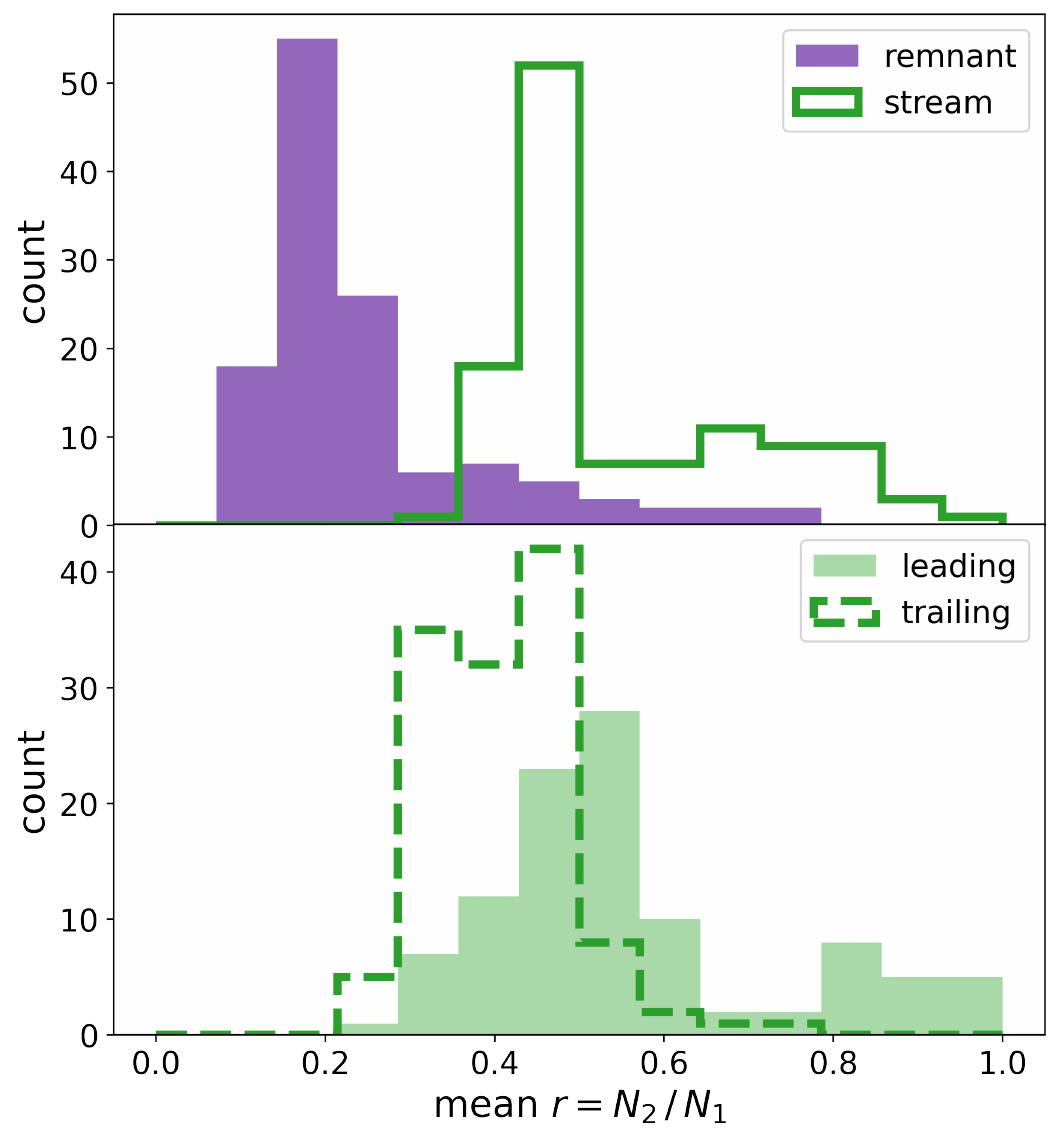} 
    \caption{The mean value of $r$ for the stream and remnant for 63 snapshots, from 21 of the most relevant simulations. Here, $r$ is the ratio of particles in the smaller satellite $N_2$ to the larger satellite $N_1$. In the top panel, we show that the smaller satellite's contribution to the stream is very often much greater than its contribution to the remnant. In the bottom panel, we show that the stream distribution is not evenly split between the leading tail and the trailing tail.}
    \label{fig:ensemble_ratios}
\end{figure}

To confirm that this stream-dominated scenario is a common occurrence we looked for more examples in our simulation suite where, after at least 4 Gyr of pre-infall mixing and several Gyr into the post-infall merger (4 -- 6 Gyr), the ratio $r = N_2 / N_1$ was higher outside the central remnant. Therefore we use 21 simulations, and take the last three Gyr interval snapshots from these simulations, totalling 63 snapshots. We select only these 21 simulations to be certain that our sample is sufficiently mixed in the pre-infall merger to be seen as a single formed system and not two distinct systems still undergoing a merger. In the top panel of Fig.~\ref{fig:ensemble_ratios} we plot the mean value of $r$ between the remnant and stream for these 63 snapshots. It is clear from Fig.~\ref{fig:ensemble_ratios} that the mean value of $r$ peaks at around $r=0.2$ for the remnant and $r=0.5$ for the stream. In the bottom panel of Fig.~\ref{fig:ensemble_ratios}, we show that the distribution of mean $r$ is not the same for the leading and the trailing arms across our simulations. This appears to challenge our expectation that the mass distribution for a disrupted satellite would be the same in both tails \citep[e.g.][]{niederste-ostholt2012tale}. Therefore, it appears that this double-system model may help in explaining the metallicity [M/H] asymmetry found between the leading and trailing tails, as shown by \citet[][]{cunningham2023chemical}. Given that our simulations do not match the exact observed properties of Sgr, we do not want to make any strong quantitative claims.

\section{Observational Data}\label{sec:data}

\begin{table}
    \centering
	\caption{Number of member stars $N$, median [Mg/Fe], and median [Mg/Fe] error values for the Sgr stream and Sgr remnant.}
    \label{tab:nmembers}
	\begin{tabular}{llll} 
	     \hline
		 System & N & Median [Mg/Fe] & Median [Mg/Fe] error\\
      \hline
      Sgr Remnant & 886 & -0.06 & 0.03\\
      Sgr Stream & 224 & -0.04 & 0.02\\
	\hline
	\end{tabular}
\end{table}

\begin{figure*}
    \centering
    \includegraphics[width=0.8\textwidth]{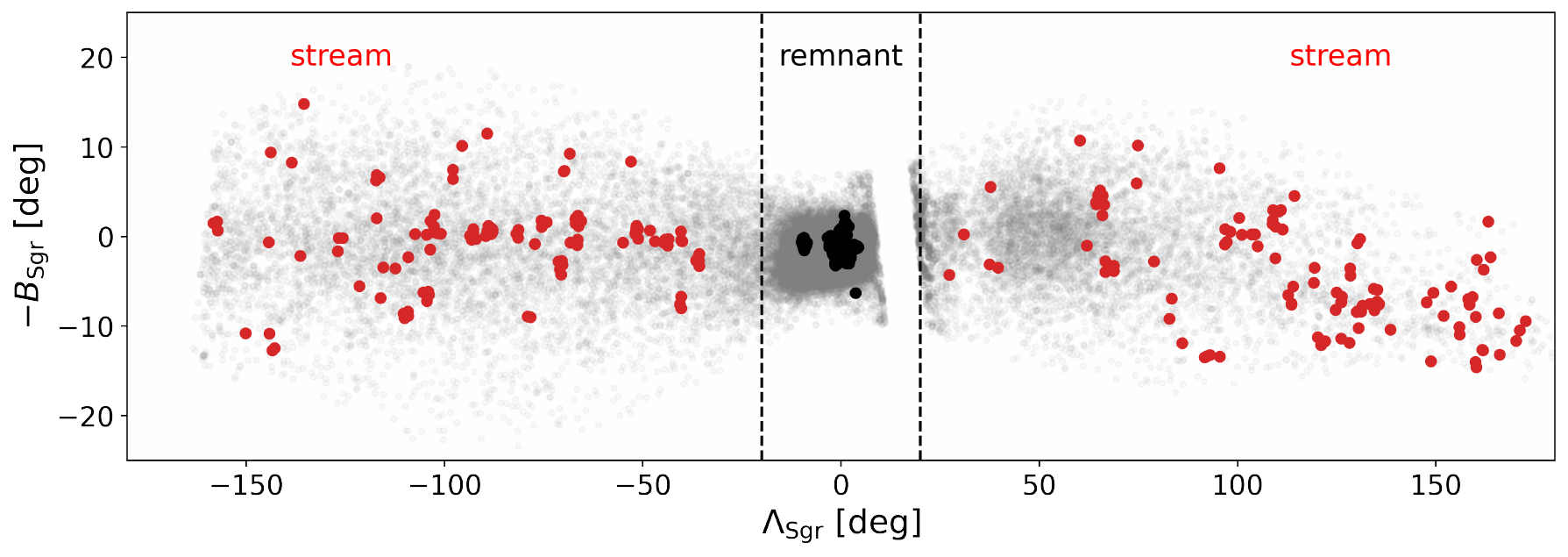}
    \caption{The stars for which APOGEE measures both [Mg/Fe] and [Fe/H] within the Sagittarius stream, presented in Sagittarius orientated stream coordinates $(\Lambda_{\rm Sgr}, B_{\rm Sgr})$. To separate the stream from the remnant, we make a cut in $\Lambda_{\rm Sgr}$, shown by the vertical dashed black lines. The colours of red for stream and black for remnant are consistent with every subsequent figure. The faded grey points show the other particles in the stream, for which we do not have chemical abundance values.}
    \label{fig:all_stream}
\end{figure*}

\begin{figure}
    \centering
    \includegraphics[width=0.9\columnwidth]{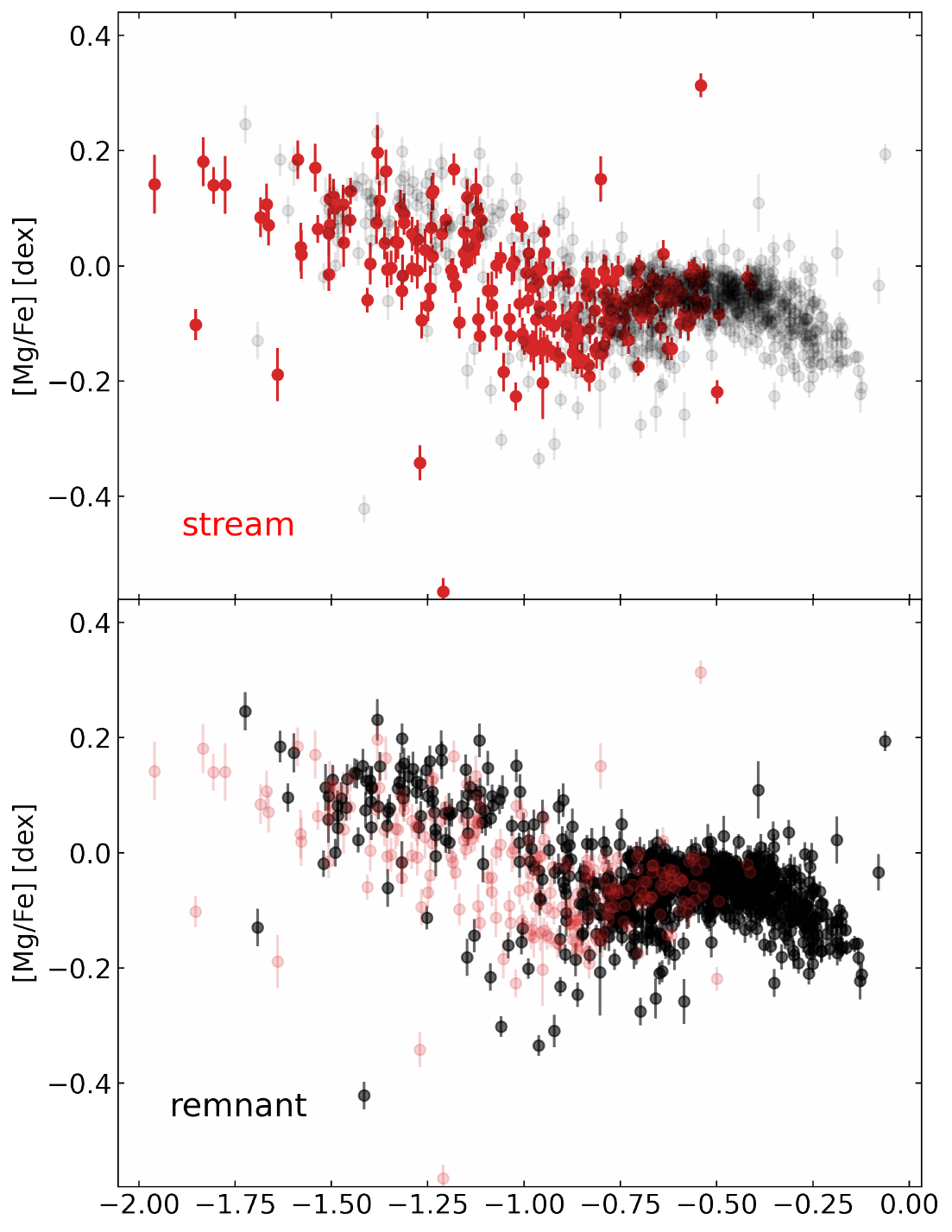}
    \caption{Chemical data from APOGEE of stars in Sagittarius. The top panel highlights the stream stars in red, with the remnant stars in faded black. The bottom panel highlights the remnant stars in black, with the stream stars in faded red. This presentation of the data makes the offset of the stream and remnant stars more clear, which is especially visible for [Fe/H] $\lesssim -1.0$ dex.}
    \label{fig:data_separated}
\end{figure}

\begin{figure}
    \centering    \includegraphics[width=0.9\columnwidth]{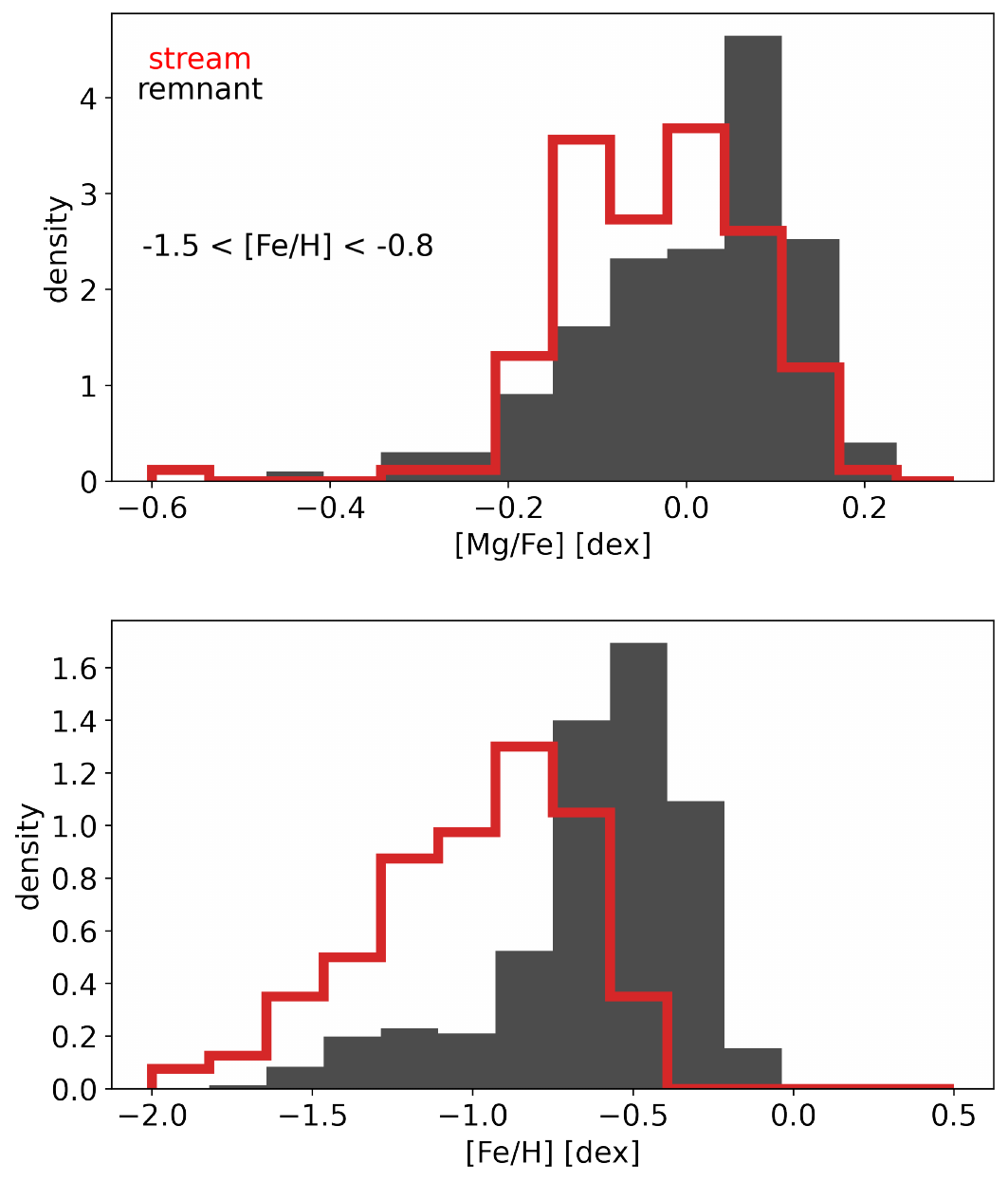}
    \caption{Distributions of [Mg/Fe] and [Fe/H] in Sagittarius, separated by stream (red line histogram) and remnant (dark filled histogram). We show the distribution of [Mg/Fe] for [Fe/H] ranges where there is sufficient data from both the stream and the remnant i.e. -1.5 < [Fe/H] < -0.8}
    \label{fig:1dchem}
\end{figure}

In this work, we study the chemical structure of Sgr and its stream using a catalogue of 1100 Red Giant Branch (RGB) stars. The Sgr members were found by \citet[][]{vasiliev2021tango}, and all chemical abundances are obtained from the 17th data release of APOGEE \citep[][]{APOGEEDR17}. In this section, we describe the coordinate system used and give an overview of our chemical catalogue and data binning.

\subsection{Coordinate system}

As in \citet[][]{vasiliev2021tango} and \citet[][]{cunningham2023chemical}, the Sgr-stream coordinate system $(\Lambda_{\rm Sgr}, B_{\rm Sgr})$ is defined so that $\Lambda_{\rm Sgr}$ increases towards the leading arm, and is centred on $(\Lambda_{\rm Sgr}, B_{\rm Sgr}) = (0, 1.5)$ deg. This follows the conventions introduced by \citet[][]{majewski2003two} and \citet[][]{belokurov2014precession}, to enforce a right-handed coordinate system.

\subsection{Member selection}

The catalogue of Sgr members with chemical abundances used in this work is created by applying the same selection cuts as were made in \citet[][]{vasiliev2021tango} to the APOGEE DR17 data \citep[][]{APOGEEDR17}. The original \citet[][]{vasiliev2021tango} catalogue is a selection of $\sim 55,000$ Red Giant Branch (RGB) stars. The stars were obtained from  cross-matching \textit{Gaia} DR2 \citep[][]{gaia2018gaia} with 2MASS \citep[][]{skrutskie2006two}, where Sgr members were selected based on proper motions, colours and absolute magnitudes. As well as these cuts, we enforce $\log(g)~<~3$, a metallicity error cut of [Fe/H]$_{\rm error}~<~0.25$ dex, and the APOGEE flags described in \citep[][]{belokurov2022dawn}. This leaves us with $\sim1100$ RGB stars, with a median error on [Mg/Fe] of 0.02 dex. The errors on the APOGEE measurements are shown in Fig.~\ref{fig:data_separated}. We choose this data to ensure that our sample is quite pure. As stated in Sec.~2.3 of \citet[][]{vasiliev2021tango}, the contamination is likely around $3\%$. Previously, however, \citet[][]{vasiliev2021tango} obtained radial velocity measurements from various spectroscopic surveys. Therefore, with the addition of verified \textit{Gaia} DR3 radial velocities, the contamination is likely reduced further.

\subsection{Stream vs remnant}

Motivated by the simulations discussed in Sec.~\ref{sec:debris_distribution}, we believe the most fruitful way of looking for evidence of a second population in Sgr is to compare the remnant with the extended stream. 
To separate the stars into stream members and remnant members, we make a cut on Sgr-stream coordinate $\Lambda_{\rm Sgr}$. Stars with $|\Lambda_{\rm Sgr}| < 20$ are assigned as remnant members, and those with $|\Lambda_{\rm Sgr}| > 20$ are assigned as stream members. This leaves us with $N_{\rm str} = 224$ stream stars and $N_{\rm rem} = 886$ remnant stars. We summarise this in Table.~\ref{tab:nmembers}. The Sgr-stream coordinates of our sample are presented in Fig.~\ref{fig:all_stream}, but note we have flipped the axis of $B_{\rm Sgr}$ in this plot following the convention in \citet[][]{vasiliev2020last}. The dashed vertical lines indicate the cuts that divide the data into stream members and remnant members. Throughout this work, the stream stars are always identified by the red, and the remnant stars by black. In Fig.~\ref{fig:data_separated}, we present a scatter plot of the [Mg/Fe]--[Fe/H] distribution of the stream stars (red) and remnant stars (black), with errors from APOGEE. We also present the 1-d distribution of [Fe/H] and [Mg/Fe] in Fig.~\ref{fig:1dchem}. 

\section{Results \& Discussion}\label{sec:results}

\begin{figure*}
    \includegraphics[width=0.95\textwidth]{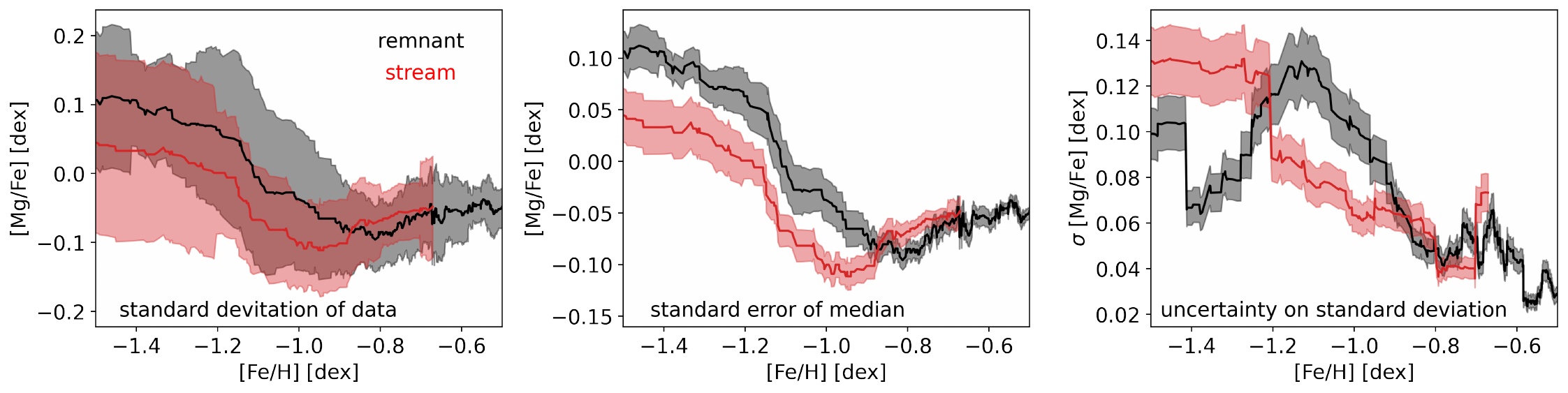}
    \caption{Moving window median, median standard error, standard deviation, and standard deviation uncertainty of [Mg/Fe] in Sagittarius, for a fixed number moving window of size $N=40$. In this figure, we highlight the lower metallicity portion of the data where there is sufficient data to have meaningful statistics i.e. $-1.5 <$ [Fe/H] $\lesssim -0.5$ dex. As usual, the stream values are shown in red and the remnant values are shown in black. In the bottom left of each panel we describe what the shaded error ranges represent. \textit{Left:} the bold lines show the moving window medians, with the surrounding faded area showing the standard deviation in the data. \textit{Middle:} the bold lines show the moving window medians, with the surrounding faded area showing the standard error on the median (i.e. $\sigma\sqrt{\pi / 2N}$). \textit{Right:} the bold lines show the moving window standard deviations, with the surrounding faded area showing the uncertainty on the standard deviation (i.e. $\sigma / \sqrt{2(N-1)}$).}
    \label{fig:track_3plot}
\end{figure*}

In this section we take a more detailed look into the chemical data from the Sgr stream and remnant. To investigate and quantify the difference between the two population, we look at their respective [Mg/Fe]--[Fe/H] median tracks, dispersion, and what this scenario could imply for the star formation history of Sgr.

\subsection{Separation of tracks}

The first step in measuring the difference between the two chemical distributions (stream and remnant) is simply finding the median track in [Mg/Fe] -- [Fe/H] space and visually inspecting them. In doing this, we hope to find significant non-overlap of the two tracks, which would point to unique star formation histories. In this work, we calculate the moving window median of [Mg/Fe] values by two methods. The first method involves using a moving window where the number of particles in each window calculation is fixed -- \textit{fixed number} window. For the second method, we use a moving window of fixed size in dex -- \textit{fixed width} window. In the latter scenario the number of particles in each window is likely to be different. Since a moving window with a fixed number of data points per window will miss out some of the data, we start the moving window at the high [Fe/H] end of the data, as this is where the two tracks are known to overlap. Note also from Fig.~\ref{fig:data_separated} that there are very few stars below [Fe/H] $< -1.5$ for both the remnant and the stream. Therefore in this section we do not make any statements comparing the data below this point.

We examined the median, standard error of the median, standard deviation, and uncertainty on the standard deviation, using three different fixed number windows of $N=\{20, 30, 40\}$. We found little difference between the three cases, so present only the $N=40$ case here (in Fig.~\ref{fig:track_3plot}). This figure presents a quite striking separation of the stream and remnant track. In the leftmost panel, the bold lines indicate the median of the track, whereas the shaded region shows the standard standard deviation of the data $\sigma$. In the middle panel, the bold lines again indicate the median, but the shaded region now shows the standard error of the median ($\sigma\sqrt{\pi/2N}$) \citep[e.g.][]{williams2001weighing}. In the rightmost panel, the bold lines are the standard deviation and the shaded region shows the uncertainty on the standard deviation $(\sigma / \sqrt{2(N-1)})$ \citep[e.g.][]{benhamou2018few}. In the middle panel we see that the tracks deviate quite significantly at [Fe/H] $< -0.9$ dex, and there is very little-to-no overlap within the median standard error. In some ranges of [Fe/H], the separation is higher than $0.05$ dex.

\subsection{Dispersion of tracks}

As well as the separation of the median tracks, the existence of a second stellar population would also have implications for the dispersion of the Mg in Sgr. Given our two system model where NGC~2419 is originally part of a small dGal, we may expect a wider dispersion in the track than in the remnant. Similarly, we may expect a wider Mg distribution in total Sgr system than in other similar dGals. As observed in small dGals, the NGC~2419 progenitor may exhibit high chemical dispersion resulting from single chemical enrichment events and inefficient mixing \citep[e.g. see][]{venn2012nucleosynthesis, ji2016complete}. Moreover, simply having multiple star formation episodes, resulting from the presence of two systems, could widen the spread of [Mg/Fe] for a given [Fe/H] range. To explore this scenario, in the rightmost panel of Fig.~\ref{fig:track_3plot} we more clearly present the standard deviations of the remnant and stream along with their uncertainties. It is clear to see that the standard deviations of the stream becomes much higher than the remnant for [Fe/H] $< -1.2$, despite being much lower at higher metallicities. In addition to comparing the stream and the remnant, comparisons were made with other dGals (Fornax, the LMC, the SMC and Sculptor) using data from \citet{hasselquist2021} and \citet{fernandes2023}. We did not find that Sgr showed a larger dispersion in [Mg/Fe], though perhaps this is not surprising given the diversity of physical properties and star formation histories traced by these galaxies.

\subsection{Kolmogorov-Smirnov test}

\begin{figure}
    \centering
    \includegraphics[width=0.9\columnwidth]{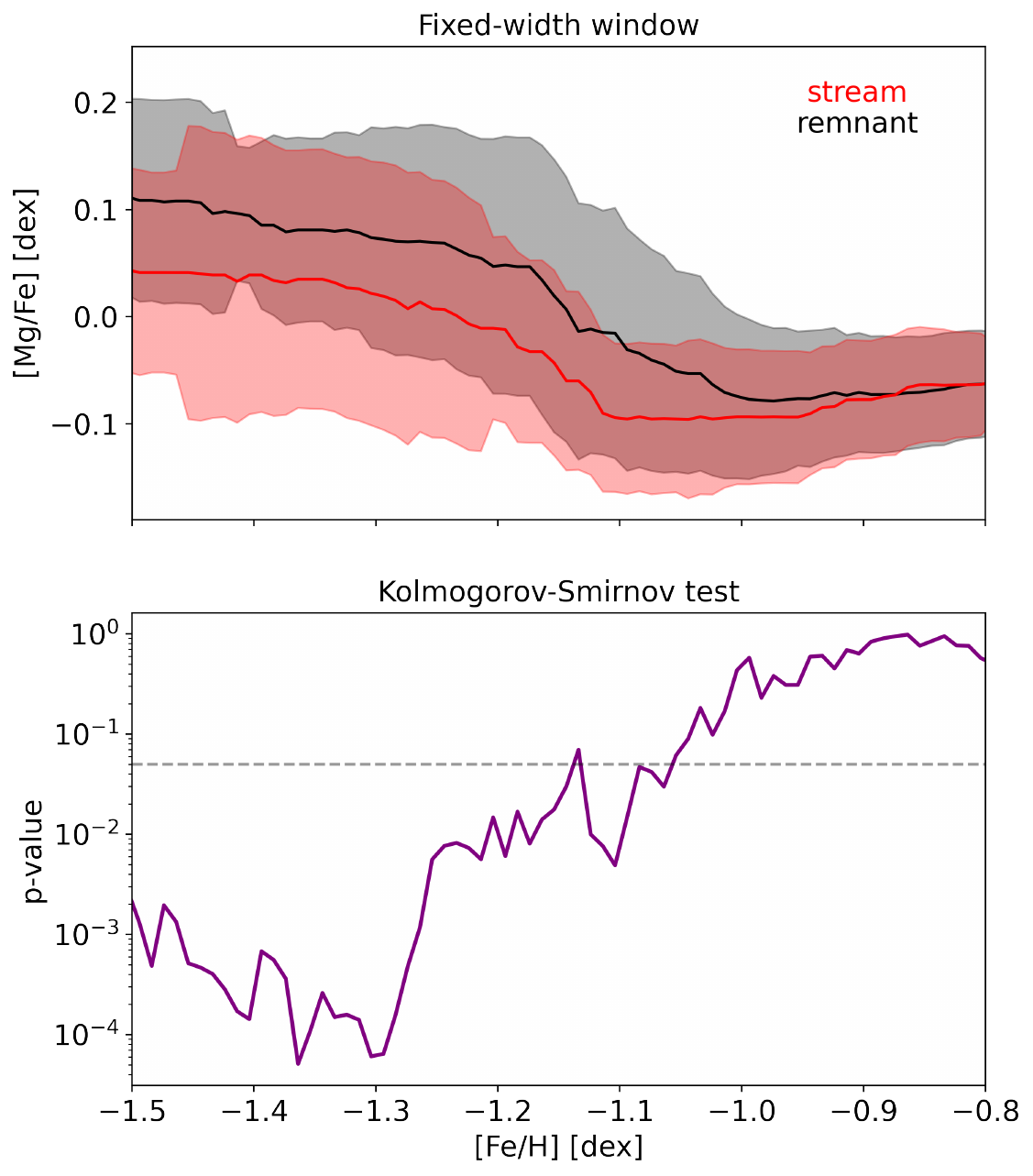}
    \caption{Moving window two-sample Kolmogorov-Smirnov test comparing the distributions of stream and remnant data at a given [Fe/H] range. \textit{Top:} Median tracks of the stream (red) and remnant (black), using a fixed dex window size of $w=0.25$ dex, as opposed to fixed particle number windows. \textit{Bottom:} Resulting p-values. The horizontal dashed line indicates a p-value of 0.05. We can see that there are large [Fe/H] ranges which the two samples of data fall below this value in each window size, most notably at [Fe/H] $ < -1.2$ dex.}
    \label{fig:ks_test}
\end{figure}

To more concretely investigate the separation of the two data sets, we use a fixed-width moving window two-sample Kolmogorov-Smirnov (K-S) test. This allows us to find the probability that, within a fixed range of [Fe/H], the two distributions were drawn from the same sample. In Fig.~\ref{fig:ks_test}, we present the result of conducting this K-S using a window of $w=0.25$ dex. We conducted this test with other sensible window sizes and found little difference from the case presented here. The top row shows the median and standard deviation in the data for this window size, while the bottom row shows the p-values from the K-S test. As always, the stream data is presented in red, while the remnant data is presented in black. This figure further reinforces that the there is significant deviation between the two distributions at [Fe/H] $< -1.1$ dex. The K-S test returns p-values below 0.05 for a wide range of [Fe/H] for the chosen size, as shown in the bottom row of Fig.~\ref{fig:ks_test}, as well as all other sensibly chosen window sizes.

\subsection{Chemical evolution}

\begin{table*}
    \centering
	\caption{Input parameters used in \texttt{flexCE} to construct the GCE model for the Sgr remnant and stream. Note that the values for the remnant were adopted directly from the best-fit model of \citet{hasselquist2021}.}
    \label{tab:gceparams}
	\begin{tabular}{lllllllll} 
	     \hline
	   Sys. & Total Gas Mass & Inflow Mass Scale & Inflow Timescale & SFE & Outflow Strength & Burst Time & Burst Duration & Burst Strength   \\
        & M$_{0}$ (M$_{\odot}$) &  M${\mathrm{i}}$ (M$_{\odot}$) & $\tau_{\mathrm{i}}$ (Gyr) & Gyr$^{-1}$ & $\eta_{\mathrm{wind}}$ & $\tau{\mathrm{b}}$ (Gyr) & $\sigma_{\mathrm{b}}$ (Gyr) & F$_{\mathrm{b}}$ \\
      \hline
      Sgr Rem. & $3\times10^{9}$ & $6\times10^{10}$ & 1 & 0.01 & 17.5 & 5 & 0.5 & 0.01\\
      Sgr Stream & $3\times10^{9}$ & $6\times10^{10}$ & 1 & 0.005 & 17.5 & 7.0 & 1.5 & 0.01 \\
	\hline
	\end{tabular}
\end{table*}

\begin{figure}
    \centering
    \includegraphics[width=0.9\columnwidth]{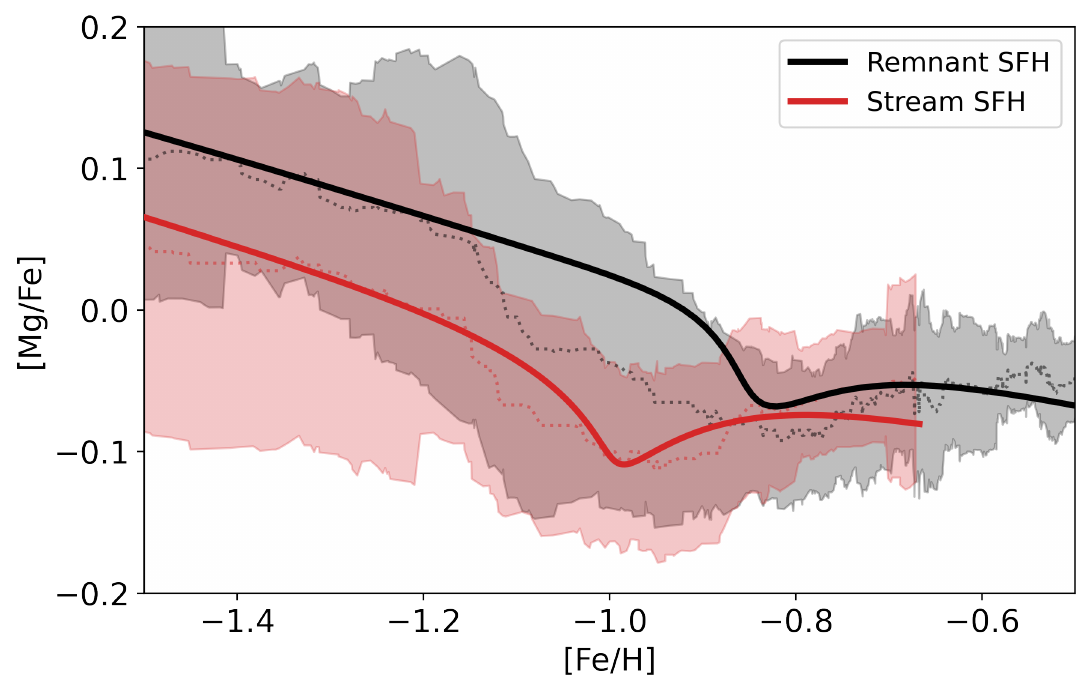}
    \caption{Galactic chemical evolution (GCE) fit to the (fixed number, $N=40$) moving window median tracks for the Sgr stream (red) and remnant (black). The bold solid lines show the GCE fits, the faint dotted lines shown the median tracks (the same as the leftmost panel of Fig.~\ref{fig:track_3plot}), and the shaded region is the associated standard deviations. The fits are calculated using \texttt{flexCE}, with the parameters in Table.~\ref{tab:gceparams}.}
    \label{fig:sfh}
\end{figure}

Given the apparent separation of the stream and remnant tracks, it is interesting to fit chemical evolution models to them. Although a dedicated exploration of the star formation history (SFH) of the stream is beyond the scope of this work, we perform an initial and demonstrative fit with a galactic chemical evolution (GCE) model. To do this, we use the \textsc{python}-based GCE code, \texttt{flexCE} described in \citet{flexce}. Briefly, \texttt{flexCE} is a one-zone chemical evolution model that accounts for global galactic physics (inflows and outflows), properties of star formation (the initial mass function and star formation efficiency) and feedback (supernovae delay time distributions and stellar yields). We adopt the same variations to the \texttt{flexCE} code as H21, making star formation efficiency a function of time using the description of \citet{nidever2020}. This formulation facilitates the modelling of star formation bursts in the GCE (see the Appendix of H21 for more information).

To explore the GCE of our stream population, we begin with the best-fit GCE model for Sgr from H21 parameterised in their Table.~\ref{tab:gceparams}. As in H21, after modifying \texttt{flexCE}, we parameterise our GCE models using eight input parameters. These values are listed in Table~\ref{tab:gceparams}. In the case of the remnant, we take the values straight from H21 with the exception of the SFE. We find that the SFE must be three times smaller than the H21 value to fit out remnant in [Mg/Fe] vs. [Fe/H]. To create a mock GCE for the stream population, we work from the Sgr remnant model, making adjustments only to the SFE of the model to reflect a smaller, less efficient SF progenitor, the time at which the star burst begins in the galaxy ($\tau_{\mathrm{b}}$) and the duration ($\sigma_{\mathrm{b}}$) of the burst. We wander away from the H21 Sgr solution until a good fit for the [Mg/Fe] vs. [Fe/H] stream track is found. Our best-fit parameters from this exploration are listed in the second row of Table~\ref{tab:gceparams}.

The resulting fit to our remnant and stream tracks in [Mg/Fe] vs. [Fe/H] are shown in Fig.~\ref{fig:sfh}. It is important to note that the solution we find to fit the stream track may not be unique and is instead only demonstrative. This is highlighted in Fig.~3 of \citet{flexce}, where it can be seen that some of the \texttt{flexCE} input parameters can effect similar changes on the track of [$\alpha$/Fe] vs. [Fe/H]. Bearing this in mind, it is interesting that the burst times for both remnant and stream are nearly temporally consistent (within $\sim1.5$~Gyr of each other, after accounting for burst duration). Another interesting thing to note is that the slower SFE assumed for the stream under the presumption of a smaller progenitor, is consistent with the [Mg/Fe] stream track. Lastly, given that our hypothesis states that both the stream and the remnant contain a mixture of two populations, our GCE fits are biased representations of the larger and smaller dGals' tracks. That is, given a mix of two populations, the smaller dGal's track should be lower than our model, and the larger dGal's track should be higher. In the future, we will perform a more thorough GCE analysis of both the stream and remnant, exploring for example, models which involve SF via a mixture of gasses from both progenitors.

\section{Conclusions}\label{sec:conclusions}

At the heart of the field of \textit{Galactic Archaeology} is the intent to use the chemical content of stellar populations to unravel their dynamical origins. We apply this principle to the dwarf spheroidal galaxy Sagittarius (Sgr); the unusual morphology of Sgr gives rise to exciting hypotheses as to its genesis, upon which its chemistry may shed light. In a previous work, we introduced the idea that a disrupted binary dwarf galaxy (dGal) merger may explain the bifurcation in the Sgr stream. Moreover, this situation may elucidate the nature of the globular cluster NGC 2419, which shows signs of being an isolated nuclear star cluster. Motivated by simulations that suggest a second smaller dGal's stars would be found mostly in the stream, we separated Sgr into two parts: a) the extended tidal stream, b) the remnant core. We do this by making a cut on stream coordinate $\Lambda_{\rm Sgr}$. We examine the $\alpha$ element content of these two components, using Mg as a proxy, and find differences in a few ways.

The median [Mg/Fe] -- [Fe/H] track of the Sgr stream is shown to be consistently below the track of the remnant for $-1.5 <$ [Fe/H] $<-0.9$ dex. Moreover, the standard deviation of the stream track seems larger than the remnant for about the range of $-1.5<$ [Fe/H] $<-1.2$. Using a fixed width moving window Kolmogorov-Smirnov test, we find that there are significant regions for which the remnant and stream have a probability of less than 0.05 of being drawn from the same distribution. A Galactic chemical evolution fit to the two populations further shows that it is perfectly reasonably for the two population to have unique star formation histories.

Despite our findings, it is possible that differences between the stream and remnant could arise from low number statistics or a chemical gradient in the Sgr progenitor. We present our findings here to open further discussion. In future work, we wish to revisit the entire catalogue of our simulations, to more closely examine the how the smaller satellite's debris is distributed between the leader and trailing tail. If genuine asymmetry is a commonly occurring feature of such simulations, this may help us understand the observed metallicity asymmetry in the Sgr stream. We must also tailor our simulations to much closely match the orbital properties of Sgr before we make any concrete claims.

\section*{Acknowledgements}

The authors would like to thank Kim Venn for useful discussion, in addition to the Cambridge Streams group. EYD thanks the Science and Technology Facilities Council (STFC) for a PhD studentship (UKRI grant number 2605433). SM and VB acknowledge support from the Leverhulme Research Project Grant RPG-2021-205: "The Faint Universe Made Visible with Machine Learning".

\section*{Data Availability}

The data we have made use of is all publicly available from APOGEE DR17, and can be reproduced using the description provided in Sec.~\ref{sec:data}. The simulations in this project can be reproduced with publicly available software using the description provided in Sec.~\ref{sec:simulations}.



\bibliographystyle{mnras}
\bibliography{contamination} 




\appendix


\bsp	
\label{lastpage}
\end{document}